# Origin of Multifractality in Solar Wind Turbulence: the Role of Current Sheets

Leonardo F. Gomes,[1]⋆ Tiago F. P. Gomes,[1] Erico L. Rempel[1,2] and Sílvio Gama[3]
[1]*Aeronautics Institute of Technology (ITA), 12228-900, São José dos Campos, SP, Brazil*
[2]*National Institute for Space Research (INPE), P. O. Box 515, 12227-010, São José dos Campos, SP, Brazil*
[3]*Mathematics Center of the Porto University (CMUP), Mathematics Department, Faculty of Sciences, University of Porto, R. Campo Alegre s/n, 4169-007 Porto, Portugal*

2 December 2022

**ABSTRACT**
In this work, a multifractal framework is proposed to investigate the effects of current sheets in solar wind turbulence. By using multifractal detrended fluctuation analysis coupled with surrogate methods and volatility, two solar wind magnetic field time series are investigated, one with current sheets and one without current sheets. Despite the lack of extreme-events intermittent bursts in the current sheet-free series, both series are shown to be strongly multifractal, although the current sheet-free series displays an almost linear behavior for the scaling exponent of structure functions. Long-range correlations are shown to be the main source of multifractality for the series without current sheets, while a combination of heavy-tail distribution and nonlinear correlations are responsible for multifractality in the series with current sheets. The multifractality in both time series is formally shown to be associated with an energy-cascade process using the $p$-model.

**Key words:** multifractals – turbulence – data analysis – statistical – solar wind

## 1 INTRODUCTION

Fractals have been widely employed in nonlinear analysis along the past decades as a form of representing the complex topological structures produced by dynamical systems. These topological structures are subsets of the phase space that may represent chaotic attractors, stable or unstable manifolds, boundaries between basins of attraction, etc. Thus, when dynamical systems are investigated through nonlinear time series analysis, the fractal indices computed from the time series somehow represent the complexity of the structure of an underlying set on which the solution lies. Additionally, the dynamical structure could be represented either by a monofractal or a multifractal process. A monofractal process has a scaling law for a fluctuation function which is a linear function of statistical moments with a single scaling exponent. A multifractal process has a power-law scaling which is a nonlinear function of statistical moments with a range of scaling exponents (Salat et al. 2017). A monofractal scaling is to be expected from dynamical processes behind perfectly self-similar fractal sets, like deterministically generated Cantor sets (Cantor 1883), or even from white noise time series (Ihlen 2012); multifractals, on the other hand, are observed in inhomogeneous systems, such as strongly intermittent turbulence, where the presence of strong fluctuations related to coherent structures localized in space generate a departure from Gaussianity in probability distribution functions (PDFs) of small-scale structure functions (Carbone et al. 2004), as seen in several analyses of observational magnetohydrodynamic data (see, e.g., Marsch & Tu (1998), Burlaga (2001), and Bruno (2019) for reviews on turbulence, intermittency and multifractal scalings in the solar wind).

A series of recent works have confirmed the complex and multifractal nature of solar wind fluctuations. Chang et al. (2004) studied the origin of complexity in space plasmas using MHD simulations, dynamic renormalization group and wavelet analysis, arguing that the turbulent plasmas in the solar wind and auroral regions are dominated by a combination of propagating modes and nonpropagating intermittent nonlinear structures, whose interactions with charged particles may lead to the energization of plasma populations such as auroral ions. Macek (2007) employed Voyager magnetic field data in the outer heliosphere and Helios plasma data in the inner heliosphere to show that multifractal spectra of intermittent solar wind fluctuations are consistent with that of the generalized two-scale weighted Cantor set. Bolzan & Rosa (2012) analyzed magnetic field data from the ACE satellite and conjectured that the presence of large scale coherent structures during coronal mass ejections (CME) decreases the multifractality, when compared with periods after the CME events. Wavelet-leader multifractal analysis of magnetospheric dissipations, as measured by the AL index, reveal that the magnetosphere is a multi-scale, complex, turbulent system, driven into a non-equilibrium self-organized state, which may explain the observations of repeatable and coherent substorm phenomena with underlying complex multifractal behavior in the plasma sheet (Valdivia et al. 2013). The interaction of the solar wind with the Earth's magnetosphere also contributes for multifractality in measurements of the geomagnetic activity, such as the geomagnetic induced current (Wirsing & Mili 2020) and the Dst index (Ogunjo et al. 2021), although internal sources of multifractality must also be considered, as Gopinath (2016) suggests that multifractality of the auroral electrojet index is fairly independent of the solar activity cycle. Wawrzaszek et al. (2019) characterized multifractality in intermittent turbulence of heliospheric magnetic field fluctuations from Ulysses spacecraft, concluding that intermittency/multifractality decreases with heliospheric distance, a result that was confirmed by Kiran et al. (2021). Recent analysis of electron density fluctuations in the E-F valley region of the ionosphere performed with the multifractal detrended fluctuation analysis (MF-DFA) method show that irregularities are multifractal, asymmetric, intermittent and non–homogeneous (Neelakshi et al. 2022).

⋆ E-mail: leofgb@ita.br

The direct link between intermittency and multifractality of magnetic and velocity field fluctuations in the solar wind was made clear in Salem et al. (2009). Using data from the Wind spacecraft, they applied the Haar wavelet transform to filter out intermittency from the time series and showed that the scaling exponents for the structure functions behave as a linear function of statistical moments, as in monofractal processes, therefore attributing multifractality in the solar wind to intermittency. Gomes et al. (2019) obtained a similar linear scaling after filtering out the current sheets from Cluster-1 intermittent magnetic field data, suggesting that the current sheets are the coherent structures responsible for the nonlinear scaling of the structure functions in the solar wind. This was confirmed after inspection of time series of days when current sheets were absent, that also showed a linear scaling.

A question remained on whether the linear scalings found by Salem et al. (2009) and Gomes et al. (2019) indeed imply that the filtered time series are monofractal or not, i.e., is the nonlinearity of the distribution of scaling exponents of structure functions a general measure of multifractality or is it just an indication of intermittency, one among different possible sources of multifractality? One of the goals of the current work is to answer this question. In this sense, it is important to stress that the origin of multifractality is not always related to fat-tailed PDFs, as it may also be caused by different correlations in small and large fluctuations, such as linear or nonlinear correlations (Kantelhardt et al. 2002; Wu et al. 2018). The source of multifractality can be investigated by producing surrogates from the original time series. Two types of surrogates are useful in this context (Theiler et al. 1992; Lancaster et al. 2018). First, shuffling the amplitudes of the original signal breaks all long-range correlations, while keeping the PDF unchanged. Therefore, if the multifractality is due to fat-tailed PDFs, it cannot be removed by shuffling the series. If it is due, solely, to time correlations, the corresponding shuffled series will be monofractal. If both fat-tailed PDF and linear/nonlinear correlations are present, the multifractality of the shuffled series should be smaller than that of the original series (Barunik et al. 2012). The second type of surrogate is produced by randomizing the phases of the Fourier modes of the original time series, producing a new series with Gaussian PDF, but preserving the linear correlations of the original series. If the random phases time series becomes monofractal, then nonlinear correlations and/or non-Gaussian PDFs are the source of multifractality. If the multifractality is preserved in the random phases time series, then linear correlations are its source.

Studies of surrogate time series have been conducted to probe the origin of multifractality in a wide range of contexts, including financial markets (Barunik et al. 2012), human gate diseases (Dutta et al. 2013), near-fault earthquake ground motions (Yang et al. 2015), solar irradiance fluctuations (Madanchi et al. 2017), air pollutants (Dong et al. 2017), meteorological time series of air pressure, air temperature and wind speed (Gos et al. 2021) and rainfall records (Sarker & Mali 2021). The surrogate method was also employed in time series of CME linear speed during solar cycle 23 to conclude that the multifractality is due to both the broad PDF and long range time correlations (Chattopadhyay et al. 2018). In the present paper, we use the method to reveal the role of current sheets in the origin of multifractality in the solar wind. By analyzing two qualitatively different magnetic field time series from Cluster-1, one filled with current sheets and another one void of current sheets, we develop a nonlinear methodology based on the MF-DFA method coupled with the volatility and surrogate time series. Thus, the contribution of small- and large-scale magnetic fluctuations can be quantified in different types of multifractal solar wind series. It is revealed that when the multifractality is not mainly due to the PDF, the scaling exponents display an almost linear behavior as a function of the moments of the structure function, despite the presence of strong multifractality in the series. In addition, we employ the $p$-model (Halsey et al. 1986; Meneveau & Sreenivasan 1987) to confirm that the multifractality in both types of solar wind time series can be attributed to a turbulent energy cascade process.

This paper is organized as follows. In section II, the MF-DFA methodology is briefly described; in section III, the multifractal analysis of two solar wind time series is conducted, including their volatility time series; section IV analyses the surrogate of the original and volatility time series, to determine if the source of the multifractality in the solar wind is due to PDF or correlations; section V presents the scaling exponent analysis of the original and surrogate times series; section VI describes the $p$-model analysis. Finally, section VII presents the conclusions.

## 2 MF-DFA

The MF-DFA method is a generalization of the detrended fluctuation analysis (DFA) method for quantifying long-range correlations in non-stationary time series (Kantelhardt et al. 2002). The method identifies the scaling of $q$th-order moments of the time series (Norouzzadeh et al. 2007). The MF-DFA method consists of five steps:

(i) The time series $x_k$ ($k = 1, 2, \cdots, N$) is integrated:

$$Y(i) = \sum_{k=1}^{i} [x_k - \langle x \rangle], \qquad i = 1, ..., N \qquad (1)$$

where $\langle x \rangle$ is the average value of the data set.

(ii) The series $Y(i)$ is divided into $N_s \equiv \text{int}(N/s)$ non-overlapping segments with equal lengths $s$. Since $N$ is usually not a multiple of $s$, some of the data points in the time series may be left out of the last segment. To fix this, the procedure is repeated starting from the opposite end of the time series and going backwards. Consequently, $2N_s$ segments are obtained.

(iii) The local trend for each $2N_s$ segments is calculated. Then the variance is given by

$$F^2(s, \nu) = \frac{1}{s} \sum_{i=1}^{s} \{Y[(\nu-1)s + i] - y_\nu(i)\}^2, \qquad (2)$$

for each segment indexed by $\nu = 1, \ldots, N_s$ and

$$F^2(s, \nu) = \frac{1}{s} \sum_{i=1}^{s} \{Y[N - (\nu - N_s)s + i] - y_\nu(i)\}^2 \qquad (3)$$

for $\nu = N_s + 1, \ldots, 2N_s$, where $y_\nu$ is the $m$-th degree fitting polynomial of each segment $\nu$. This polynomial detrending of order $m$ in the $Y$ profile eliminates trends up to order $m - 1$ in the original time series and specifies the type of MF-DFA applied.

(iv) The average over all segments is calculated to obtain the $q$th-order fluctuation function:

$$F_q(s) = \left\{ \frac{1}{2N_s} \sum_{\nu=1}^{2N_s} [F^2(s, \nu)]^{\frac{q}{2}} \right\}^{\frac{1}{q}}, \quad (4)$$

where, in general, the $q$ parameter can take any real value except zero. For $q = 2$, the equation returns the DFA method. Steps 2 to 4 are repeated for different time scales $s$.

(v) The scaling behavior of the fluctuation function is defined by the log-log plot of $F_q(s) \times s$ for each value of $q$. If $x_i$ have long-range correlations, for large values of $s$, $F_q(s)$ increases as a power-law,

$$F_q(s) \sim s^{h(q)}. \quad (5)$$

The scaling exponents $h(q)$ are the generalized Hurst exponents, defined as the slope of the log $F_q(s) \times \log(s)$ graph, where for $h(2)$ we have the standard Hurst Exponent (Hurst et al. 1965). For positive values of $q$, $h(q)$ describes the scaling behavior of segments with large fluctuations and for negative values of $q$, $h(q)$ describes the scaling behavior of segments with small fluctuations. For monofractal series, $h(q)$ is independent of $q$, but for multifractal series $h(q)$ depends on $q$. The generalized Hurst exponent is directly related to the Renyi exponent (Renyi 1976) $\tau(q)$ by

$$\tau(q) = q\, h(q) - 1. \quad (6)$$

Besides $h(q)$, another way to characterize the multifractality of a time series is by the singularity spectrum $f(\alpha)$, which is related to $\tau(q)$ via a Legendre transform,

$$\alpha = \tau'(q) \quad \text{and} \quad f(\alpha) = q\,\alpha - \tau(q), \quad (7)$$

where $\alpha$ is the singularity exponent. This $f(\alpha) \times \alpha$ relation represents the multifractal spectrum and has a concave parabolic shape.

From the multifractal spectrum, it is possible to obtain a set of parameters to characterize each series: (i) the $\alpha$ value where $f(\alpha)$ is maximum, $\alpha_0$; (ii) the $\alpha$ width, $\Delta \alpha = \alpha_{max} - \alpha_{min}$, where $\alpha_{min}$ and $\alpha_{max}$ are, respectively, the minimum and maximum values of $\alpha$ that mark the base of the concave parable in the multifractal spectrum ($\Delta \alpha$ is a measure of multifractal strength); (iii) the asymmetry parameter:

$$A = \frac{\alpha_{max} - \alpha_0}{\alpha_0 - \alpha_{min}}, \quad (8)$$

where $A = 1$ means the spectrum is symmetric, for $A > 1$ the spectrum is right-skewed asymmetric, and for $A < 1$ the spectrum is left-skewed asymmetric (Shimizu et al. 2002; de Freitas et al. 2016). A multifractal spectrum with a long right tail has a greater contribution from small fluctuations. By contrast, a multifractal spectrum with left asymmetry has a greater influence by local fluctuations with large values (Ihlen 2012).

Another useful multifractal parameter can be extracted from the $\tau(q) \times q$ relation. As can be seen from Eq. (6), $\tau(q)$ has a linear dependence with $q$ for monofractal series, where $h(q)$ is constant. In contrast, for multifractal series, this dependence is nonlinear. The $q$-dependency of the Renyi exponent can be quantified by the coefficient of determination, $R^2$. $R^2$ measures the proportion of the variance for a dependent variable that is predictable by an independent variable in a linear regression model (Barrett 1974). The coefficient of determination is given by:

$$R^2 = 1 - \frac{\sum_{i=1}^{n}(\tau_i - \widehat{\tau_i})^2}{\sum_{i=1}^{n}(\tau_i - \bar{\tau})^2}, \quad (9)$$

where $\tau_i = \tau(q_i)$ is the observed dependent variable, $\widehat{\tau_i}$ is the corresponding predicted value and $\bar{\tau}$ is the mean of the observed data. $R^2$ varies from 0 to 1, where in our case 1 represents a perfect fit to the linear dependence model. In other words, the measure of $R^2$ for the $\tau(q) \times q$ relation will be closer to 0 for multifractal series and closer to 1 for monofractal series.

The MF-DFA method has best results if the time series are reasonably stationary, i.e., if they have a noise like structure. As suggested by Eke et al. (2002), it is possible to determine if the time series have noise like structure by computing a monofractal detrended fluctuation analysis prior to conducting the MF-DFA analysis. Time series are noise like if their Hurst exponent $h(2)$ is between 0 and 1, and they are random walk like (nonstationary) if $h(2)$ is above 1. Ihlen (2012) suggests that time series with $h(2)$ above 1.2 should be differentiated before application of the MF-DFA analysis.

## 3 MULTIFRACTAL ANALYSIS OF SOLAR WIND DATA

We analyze solar wind magnetic field data detected with the Fluxgate Magnetometer (FGM) onboard Cluster-1, with 22 Hz sampling frequency. Two time series with 24 hours are investigated, one from 2008 March 9 and one from 2016 January 25. To reduce the computational time of the analysis, the data length has been reduced by using a decimation process. The low-pass Chebychev Type I infinite impulse response filter was used with a reduction factor $M = 10$, order 8 and $0.8/M$ cut-off frequency. This decimation process is described in Gomes et al. (2019).

After decimating the time series, we apply the MF-DFA method with four input parameters: minimum scale $s_i$, maximum scale $s_f$, order of fluctuation function $q$ and polynomial order $m$. The scale refers to multiple segment sizes of the cumulative series and varies from a minimum segment size $s_i$ to a maximum $s_f$. In this work, we use $s_i = 10$ and $s_f = N$, where $N$ is the length of the time series; $q$ varies between $-20$ and $20$ with an increment of $\Delta q = 0.25$, and $m = 3$. This choice of parameters was supported by several tests. The recommendation for large time series is to use a polynomial trend order around $m = 3$; $s_f = N$ was chosen to avoid deformations in the shape of the multifractal spectra. Meanwhile, for the $q$ parameter the use of values larger than 20 does not change the shape of the spectra significantly.

### 3.1 MF-DFA analysis of the $|B|$ time series

Figure 1 shows the solar wind magnetic field time series studied in this section for days 2008 March 9 and 2016 January 25. In the upper panel, the time series for 2008 March 9 (red) and its first order differencing (black) are shown. As it was explained in the previous section, time-differencing is necessary in this case due to the high nonstationarity of this series ($h(2) = 1.23$). Throughout the remaining of this section, only the differenced time series will be used for March 9. This time series was characterized by Gomes et al. (2019) as being permeated by large-scale current sheets. The green regions in the original time series denote current sheets found with Li's method (Li 2008). The lower panel shows the time series for 2016 January 25, which is characterized by a higher degree of stationarity and the absence of current sheets (Gomes et al. 2019).

Due to its higher stationarity ($h(2) = 0.96$), there is no need to perform a differencing in this series.

Figure 2 shows different multifractal measures of the two magnetic field time series. Figure 2(a) shows the multifractal spectra, which reveal a left asymmetry for the March 09 time series (red) and a right asymmetry for the January 25 series (blue). The left asymmetry indicates the stronger contribution to multifractality coming from large fluctuations associated with values of $q > 0$ in the intermittent time series of the current sheet-filled time series of March 09; the right asymmetry found for the current sheet-free time series of January 25 points to the greater contribution of small fluctuations to the multifractality (Ihlen 2012). The width of the spectrum can be used as a measure of the degree of multifractality of the series (Shimizu et al. 2002). Comparing both spectra, it can be seen that they have almost the same width ($\Delta\alpha \approx 0.541$ for March 9 and $\Delta\alpha \approx 0.555$ for January 25), which may be surprising, since the time series of March 9 is visibly more intermittent, with strong bursts randomly interspersed in time. In this case, the difference in multifractality can be better quantified by the Renyi exponent $\tau(q)$, shown in Fig. 2(b). It reveals a nonlinear behavior for both series, but with $R^2 \approx 0.804$ for March 9 and $R^2 \approx 0.986$ for January 25, thus, March 9 displays higher multifractality.

### 3.2 MF-DFA analysis of the volatility time series

In the previous section, the degree of multifractality, as provided by the width of the multifractal spectra, could not properly distinguish between the two time series under investigation, which is unexpected, given that the original series are not only visually very different, but one of them is known to be permeated by coherent structures (current sheets) and the other is not. This is probably because although the differenced time series of 2008 March 9 is apparently more intermittent than the series of 2016 January 25, most of the abrupt changes in $|B|$ caused by the current sheets in the March 9 series have a small amplitude and, therefore, do not produce strong bursts in the time-differenced series. Such abrupt changes in $|B|$ can be enhanced by employing the volatility, thus providing a way to investigate the role of current sheets in the multifractality. In the present section, we employ the volatility to enhance the distinct features of each series due to current sheets before repeating the MF-DFA analysis.

The magnetic volatility, $\text{vol}_{mag}$, can be calculated from the standard deviations of the log magnetic return $\Delta r_{mag}(t)$ in a moving window of length $\omega$ along $N$ sample points (Tsay 2010)

$$\Delta r_{\text{mag}}(t) = \log\left(\frac{|\mathbf{B}(t+\tau)|}{|\mathbf{B}(t)|}\right), \tag{10}$$

$$\text{vol}_{\text{mag}}(j) = \sqrt{\frac{1}{\omega-1}\sum_{i=j}^{\omega+j-1}(\Delta r_{\text{mag}}(i) - \mu(j))^2}, \tag{11}$$

where $\tau$ is a time-lag, $j = 1, \ldots, N - \omega + 1$ and $\mu(j)$ is the mean $\Delta r_{\text{mag}}$ inside the window (Gomes et al. 2019). Note that since $\Delta r_{\text{mag}}$ involves computing a time difference with lag $\tau$, there is no need to difference the original time series to remove nonstationarities prior to computation of the volatility. The $\omega$ ant $\tau$ values are estimated from the Power Spectrum Density (PSD). Figure 3(a) shows the PSD for the March 9 time series, where the inertial range is the blue region between the dashed lines. This region was chosen as the frequency interval where the slope of the fitted line is -5/3, following Kolmogorov's K41 theory (Kolmogorov 1941) for fully developed turbulence (Frisch 1995). The frequency in the middle of the inertial range marks the scale used to define both $\tau$ and $\omega$. It is also the scale used in Li's method to detect the current sheets, shown in Fig. 1. In this way, we define $\tau = \omega = 50 s$. Figure 3(b) shows the PSD for the January 25 series.

Figure 4 exhibits the volatility time series for 2008 March 9 (upper panel, red) and for 2016 January 25 (lower panel, blue) from the decimated magnetic field data. Recall that the upper series has many current sheets while the lower one has none. Note that, unlike the January 25 series, the March 9 volatility series has several extreme events. Most of these high peaks are due to the abrupt changes in the magnetic field that take place when the satellite crosses a current sheet in the solar wind, as evidenced by the coincidence between extreme events in the volatility and current sheets detected by Li's method (see Fig. 2(a),(b) in Gomes et al. (2019)). As a consequence, the multifractal spectra obtained from the volatility of both series are very different, as seen in Fig. 5(a). Now, the spectrum of the intermittent time series of March 9 is much broader than the one from January 25. The $\alpha$-width is $\Delta\alpha = 0.94134$ for March 9 and $\Delta\alpha = 0.74921$ for January 25. The volatility has enhanced the contribution of the extreme events due to current sheets, thus showing the signature of coherent structures present in the solar wind that were partially hidden in the multifractal analysis of the original time series. The Renyi exponents are shown in Fig. 5(b); once again, the curve for March 9 is more concave than for January 25, reflecting its higher level of multifractality. The coefficient of determination for the Renyi exponents is $R^2 = 0.97464$ for the volatility of March 9 and $R^2 = 0.98125$ for the volatility of January 25. It is clear that the volatility has highlighted the role of current sheets in the multifractal singularity spectrum.

## 4 MF-DFA OF SURROGATE TIME SERIES

According to Madanchi et al. (2017), there are two features in a time series that can lead to its multifractality: (i) the presence of heavy-tailed PDFs, as in highly intermittent series, and (ii) the existence of linear and non-linear correlations. In this section, we try to identify the origin of the multifractality in the solar wind by means of two surrogate time series derived from the original $|B|$ data. As mentioned in the introduction, the shuffled time series is a random permutation of the original time series in the real space that destroys all temporal correlations, while keeping the same PDF for the amplitudes of $|B|$. On the other hand, the random phases surrogate is generated from the Fourier Transform of the original $|B|$ series. A new Fourier series is generated by shuffling the phases of the Fourier modes while keeping their power spectrum (Maiwald et al. 2008). The inverse Fourier transform of this new frequency spectrum is the random phases surrogate, which keeps the power spectrum and linear autocorrelation of the original series, but has a Gaussian PDF and breaks the nonlinear correlations. After generating these two surrogates, we repeat the multifractal analysis described in the previous section; if the shuffled surrogate has a multifractal spectrum which is considerably narrower than the spectrum of the original series, it means that time correlations are an important source of multifractality in the original time series. If the random phases surrogate has a multifractal spectrum which is considerably narrower than the spectrum of the original series, it means that fat-tailed PDFs and/or nonlinear correlations are important for the multifractality. Note that both kinds of multifractality mentioned above can be simultaneously present in a time series (Norouzzadeh et al. 2007; Madanchi et al. 2017). If both the shuffled and random phases surrogates produce monofractal spectra, then nonlinear correlations (but not fat-tailed PDFs) are the source of multifractality. In the following subsections, we perform this analysis for both the $|B|$ and volatility time series of 2008 March 9 and 2016 January 25.

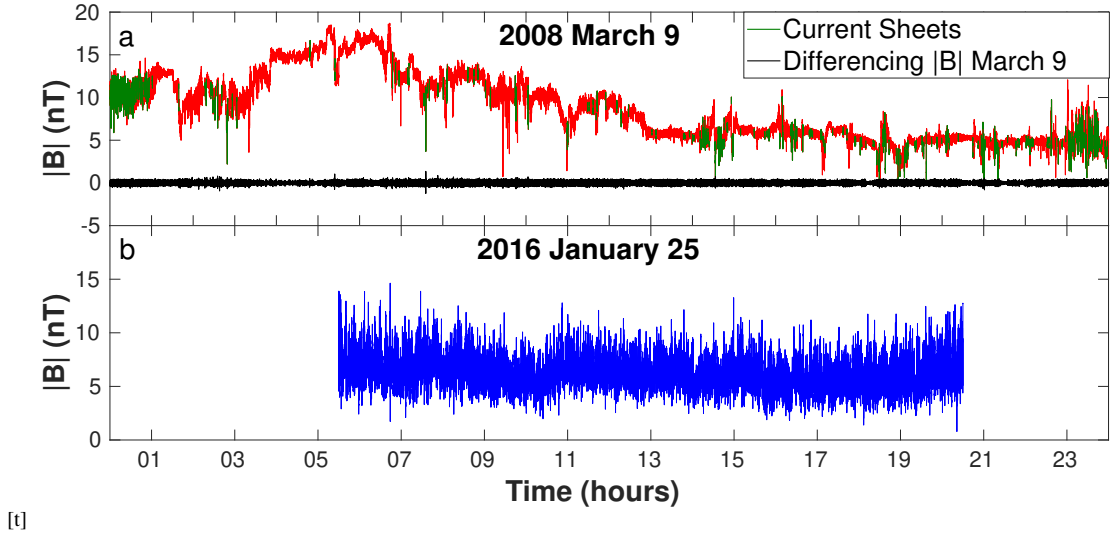

**Figure 1.** Solar wind time series of $|B|$ measured by Cluster-1. (a) For 2008 March 9 (red), containing current sheets (green), and its first order differencing (black); (b) time series of $|B|$ for 2016 January 25 (blue), without current sheets.

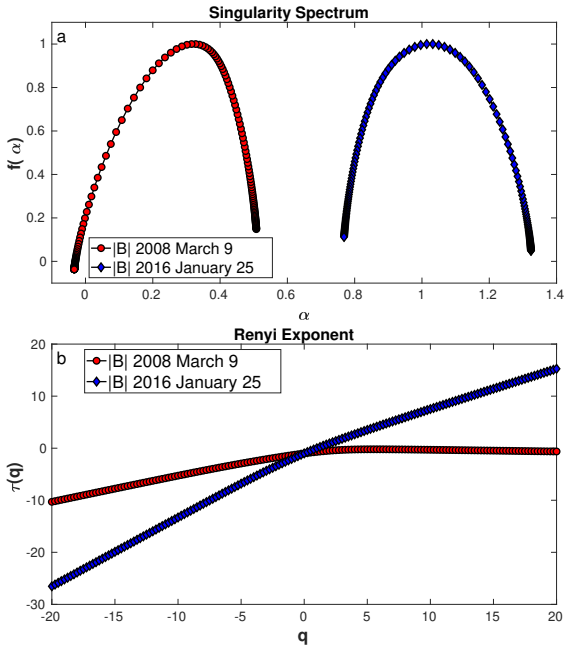

**Figure 2.** (a) Multifractal spectrum of $|B|$ for 2008 March 9 (red), and 2016 January 25 (blue). (b) Renyi exponents for 2008 March 9 (red), and 2016 January 25 (blue).

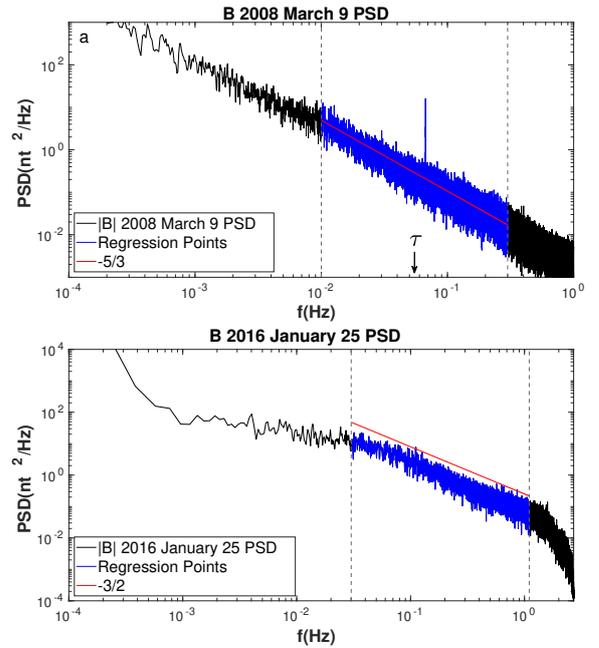

**Figure 3.** Power spectral density for solar wind magnetic field of (a) 2008 March 9, and (b) 2016 January 25. The blue region is the inertial range and the red line is the linear fit for this interval, with a slope equal to -5/3 for March 9 and slope -3/2 for January 25.

### 4.1 Magnetic Field time series, 2008 March 9

Figure 6 shows the differenced time series of $|B|$ for March 9 (red) with its shuffled (green) and random phases (magenta) surrogates. Clearly, the shuffled surrogate keeps the extreme events of the differenced $|B|$ series, but the same events are absent from the random phases surrogate.

Figure 7(a) displays the multifractal spectra for the March 9 original and surrogate time series. For the shuffled spectrum (green) we see a small reduction in the width when compared with the original one (red). This means that there is a contribution from correlations to multifractality, along with the contribution of the PDF. Considering the random phases spectrum (magenta), its width reduces drastically (the $\Delta\alpha$ variation is about 0.32), which points to a significant contribution to multifractality coming from a non-Gaussian PDF and/or nonlinear correlations. The conclusion from both spectra is that the PDF has the strongest contribution to multifractality. The contribution of the PDF is due to the presence of strong intermittent bursts (extreme events) in the March 9 time series. Since these bursts have been shown to be related to large current sheets (see Gomes et al. (2019)), the current sheets can be seen as the origin of most of the multifractality in this time series. Figure 7(b) confirms this conclusion by showing the Renyi exponent as a function of $q$, where the random phases surrogate has a smaller concavity than the shuffled surrogate.

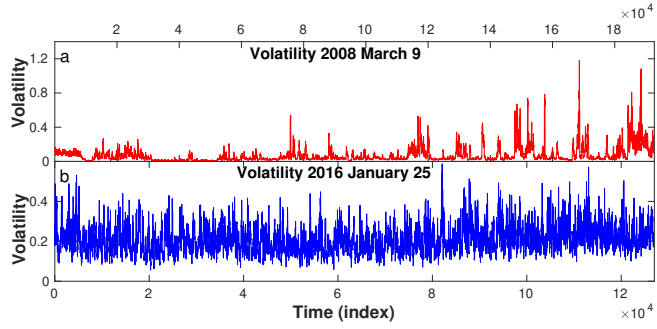

**Figure 4.** Volatility of solar wind magnetic field time series for (a) 2008 March 9, and (b) 2016 January 25.

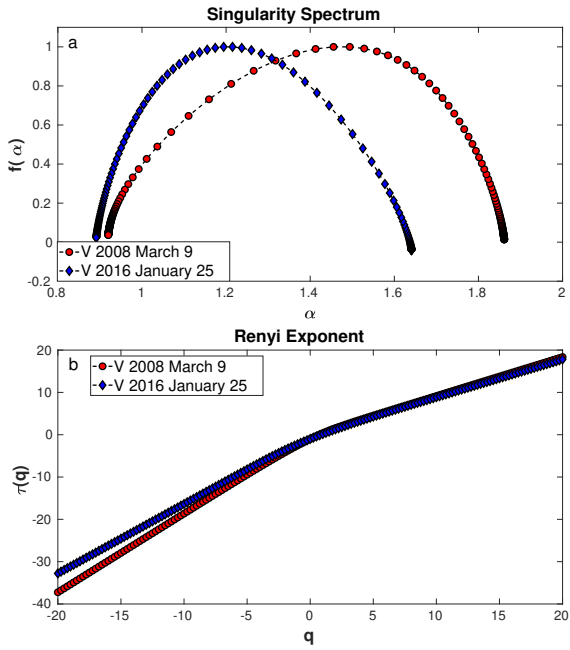

**Figure 5.** (a) Multifractal spectra for the volatility in 2008 March 9 (red), and 2016 January 25 (blue). (b) Renyi exponents for the volatility in 2008 March 9 (red), and 2016 January 25 (blue).

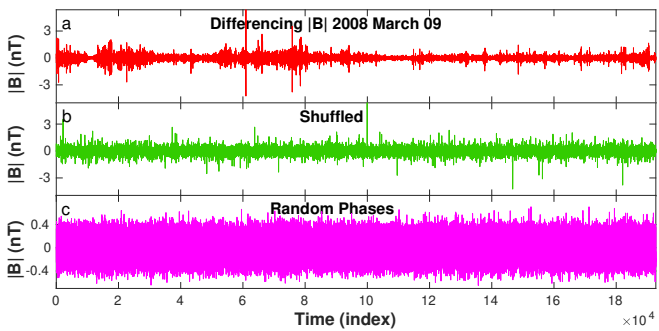

**Figure 6.** Differenced time series for 2008 March 9 (red) and the respective surrogates: shuffled (green), and random phases (magenta).

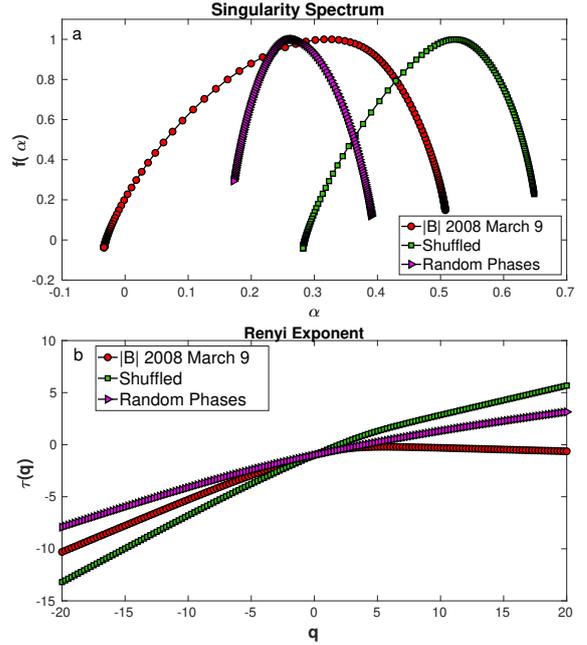

**Figure 7.** (a) Multifractal spectrum of $|B|$ for 2008 March 9 (red) and the respective surrogates: shuffled (green), and random phases (magenta). (b) Renyi exponents for 2008 March 9 (red) and the respective surrogates: shuffled (green), and random phases (magenta).

### 4.2 Magnetic field time series, 2016 January 25

Figure 8 shows the time series for January 25 (blue) with its shuffled (green) and random phases (magenta) surrogates. Figure 9(a) shows a significant width reduction in both surrogate spectra in comparison with the original volatility spectrum (blue). The spectrum of the shuffled series (green) has a width $\Delta\alpha = 0.194$, indicating a difference of 0.36 with the spectrum of $|B|$. Similarly, the spectrum for the random phases series has a small width, about $\Delta\alpha = 0.32$, a difference of 0.23 with the spectrum of $|B|$. So, there is strong influence from long-range correlations as well as non-gaussianity on the January 25 magnetic field multifractality, but the contribution of the correlations is preponderant, since the shuffled spectrum is considerably narrower than the random phases spectrum.

### 4.3 Volatility time series, 2008 March 9

We proceed with the analysis of the origin of the multifractality for March 9 using the volatility, as shown in Fig. 10 for the original (red), shuffled (green) and random phases (magenta) time series. The corresponding multifractal spectra in Fig. 11(a) show a wide parabola for the original volatility series (red) and two narrower parabolas related to its shuffled (green) and random phases (magenta) series. The random phases spectrum has a width of about $\Delta\alpha = 0.39$ and the shuffled spectrum has a width of $\Delta\alpha = 0.35$. Since both spectra have approximately the same width, it shows an important feature that was not so clear from the multifractal spectra of the $|B|$ surrogate series (Fig. 7), that is, the importance of the nonlinear correlations, which play a key role, together with the PDF, in the origin of the multifractality for the March 9 series. Since the volatility is computed with a lag-time of $\tau = 50s$, it is better suited for measuring the relevance of long-range nonlinear correlations than the time-differenced $|B|$ series. Figure 11(b) confirms that the shuffled and random phases series have almost linear Renyi exponents, thus, the series are closer to monofractal.

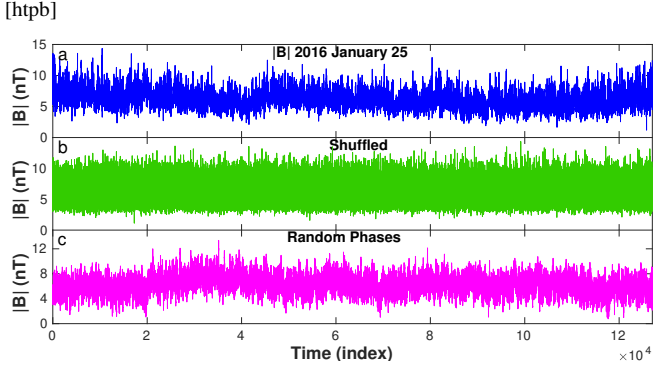

**Figure 8.** Time series for 2016 January 25 (blue) and the respective surrogates: shuffled (green), and random phases (magenta).

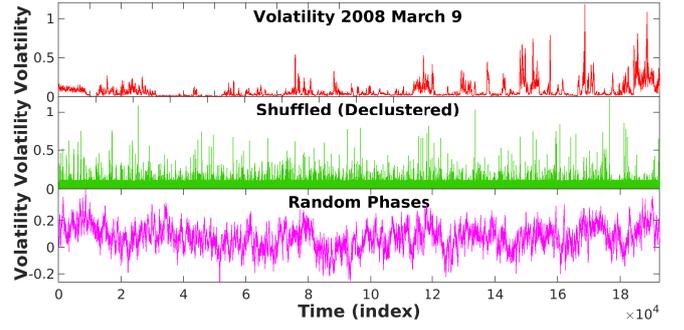

**Figure 10.** Time series Volatility for 2008 March 9 (red) and the respective surrogates: shuffled (green), and random phases (blue).

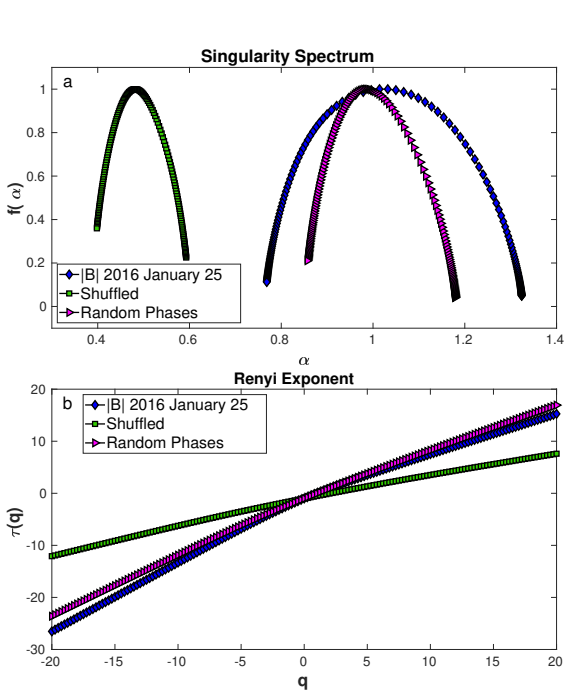

**Figure 9.** (a) Multifractal spectrum for 2016 January 25 (blue) and the respective surrogates: shuffled (green), and random phases (blue). (b) Renyi exponents for 2016 January 25 (blue) and the respective surrogates: shuffled (green), and random phases (magenta).

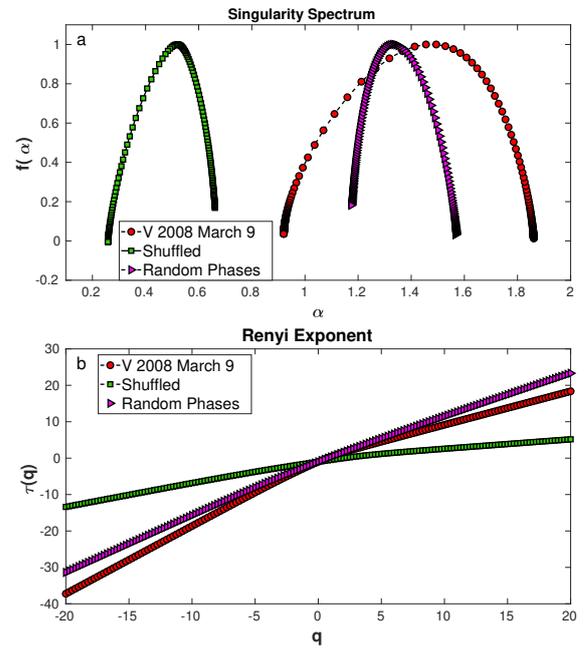

**Figure 11.** (a) Multifractal spectrum for the volatility of 2008 March 9 (red) and the respective surrogates: shuffled (green), and random phases (magenta). (b) Renyi exponents for the volatility of 2008 March 9 (red) and the respective surrogates: shuffled (green), and random phases (magenta).

### 4.4 Volatility time series, 2016 January 25

Figure 12 shows the volatility time series of the January 25 time series (blue) and its shuffled (green) and random phases (magenta) surrogates. Figure 13(a) shows the corresponding multifractal spectra. Once again, the reduction in the width for both surrogate spectra means that a mutual contribution to multifractality coming from long-range correlations and non-Gaussianity is present, with a clear predominance of the long-range correlations effects, since the shuffled spectrum is much narrower than the random phases spectrum.

A quantitative comparison of all the results for the $|B|$ time series and volatility time series of March 9 and January 25 is provided by Tables 1 to 3. Table 1 shows $R^2$ for the Renyi exponent of $|B|$ and its volatility for March 9 and January 25; Table 2 shows the width of the multifractal spectra, $\Delta\alpha$; Table 3 shows the asymmetry of the spectra, $A$. In general, all spectra for January 25 are right-asymmetric due to the importance of small scale fluctuations; for March 9, some spectra are left-asymmetric due to the importance of large-scale fluctuations, but the random phases show right asymmetry, since in the random phases surrogate the effects of non-Gaussian PDFs are destroyed.

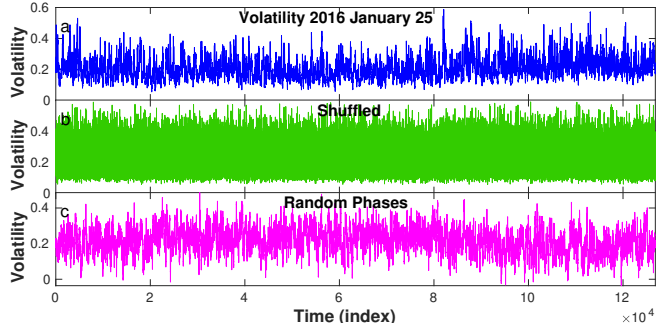

**Figure 12.** Time series of the volatility for 2016 January 25 (blue) and the respective surrogates: shuffled (green), and random phases (magenta).

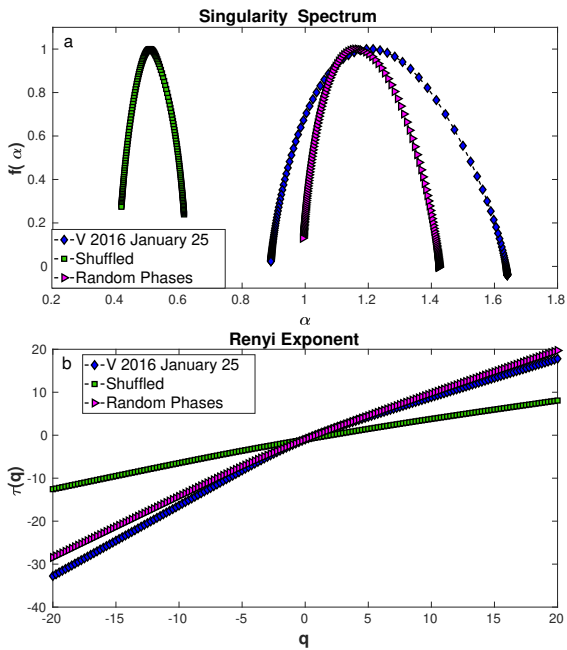

**Figure 13.** (a) Multifractal spectrum for the volatility of 2016 January 25 (blue) and the respective surrogates: shuffled (green), and random phases (magenta). (b) Renyi exponents for the volatility of 2016 January 25 (blue) and the respective surrogates: shuffled (green), and random phases (magenta).

**Table 1.** $R^2$ of the Renyi exponent for magnetic field and volatilities of 2008 March 9 and 2016 January 25

|  | March 9 |  | January 25 |  |
| --- | --- | --- | --- | --- |
|  | $|B|$ | Volatility | $|B|$ | Volatility |
| Original | 0.80413 | 0.97464 | 0.98597 | 0.98125 |
| Shuffle | 0.97505 | 0.96748 | 0.99537 | 0.99573 |
| Random Phases | 0.98185 | 0.99637 | 0.99601 | 0.99424 |

**Table 2.** Width of $\alpha$, $\Delta\alpha$, for magnetic field and volatilities of 2008 March 9 and 2016 January 25.

|  | March 9 |  | January 25 |  |
| --- | --- | --- | --- | --- |
|  | $|B|$ | Volatility | $|B|$ | Volatility |
| Original | 0.54112 | 0.94134 | 0.55568 | 0.74921 |
| Shuffle | 0.36663 | 0.40332 | 0.19468 | 0.19873 |
| Random Phases | 0.21802 | 0.39299 | 0.32181 | 0.43088 |

**Table 3.** Spectrum Asymmetry, $A$, for magnetic field and volatilities of 2008 March 9 and 2016 January 25.

|  | March 9 |  | January 25 |  |
| --- | --- | --- | --- | --- |
|  | $|B|$ | Volatility | $|B|$ | Volatility |
| Original | 0.49873 | 0.63885 | 1.10709 | 1.31215 |
| Shuffle | 0.51279 | 0.53342 | 1.30854 | 1.18283 |
| Random Phases | 1.33583 | 1.41817 | 1.45999 | 1.47002 |

## 5 ZETA FUNCTION

Another function typically employed in multifractal analyses of time series is the zeta function. Consider the structure function for $|B|$ (Frisch 1995):

$$S_p(\tau) = \langle [|B(t+\tau)| - |B(t)|]^p \rangle, \quad (12)$$

where $\langle \cdot \rangle$ is the time average, $\tau$ is the time lag and $p$ are the statistical moments for the time series of $B$. Assuming scale invariance inside the inertial range, $S_p$ follows a power law

$$S_p(\tau) \sim \tau^{\zeta(p)}, \quad (13)$$

where $\zeta(\cdot)$ is the zeta function or scaling exponent of the structure function. So, $\zeta(p)$ is obtained by the slope of the $\log S_p(\tau) \times \log \tau$ plot. The importance of this parameter comes from Kolmogorov's K41 theory (Kolmogorov 1941) and the IK (Iroshnikov-Kraichnan) theory (Iroshnikov 1964; Kraichnan 1965) of self-similarity and scale invariance inside the inertial range for a homogeneous and isotropic turbulence, where the $\zeta$ function was shown to be a linear function of $p$, with $\zeta(p) = p/3$ for K41 and $\zeta(p) = p/4$ for IK.

In Fig. 14(a), the linear K41 theoretical zeta scaling exponent function is shown by the black dashed line while the IK scaling exponent is denoted by a dotted line. The top panel (a) also shows the zeta scaling exponent computed from the time series of $|B|$ for the intermittent series of March 09 (red line with circles) and for the current sheet-free series of January 25 (blue line with diamonds). The zeta function for the March 09 series clearly departs from the linear behavior, as expected for multifractal intermittent series, but, surprisingly, the zeta function exhibits an almost linear relation with $p$ in the case of January 25, despite the fact that both series have multifractal spectra with similar widths (see Fig. 2(a)). Thus, one should be cautious before using the behavior of the scaling exponent as a definite measure of multifractality, although it is a good measure of intermittency. To confirm this result, Fig. 14(b) compares the zeta scaling exponents of the March 09 $|B|$ series (red line with circles) with the zeta scaling exponents of its random phases series (magenta line with triangles). Since the random phases series has a Gaussian PDF, it removes from the original series the intermittent extreme events responsible for the fat-tailed PDF and the zeta scaling exponent becomes linear, following the K41 line. This result con-firms the importance of the contribution from a fat-tailed PDF to the multifractality of the March 09 series. In Fig. 14(c), the same analysis is done for the January 25 series, where both the original series and its random phases show an IK linear behavior, since none of the series has fat-tailed PDF, although they have multifractal spectra (see the blue and magenta spectra in Fig. 9(a)). We conclude from this that the $\zeta$-function is a good measure of multifractality due to PDF, but misses the contribution of long-range correlations to the multifractality.

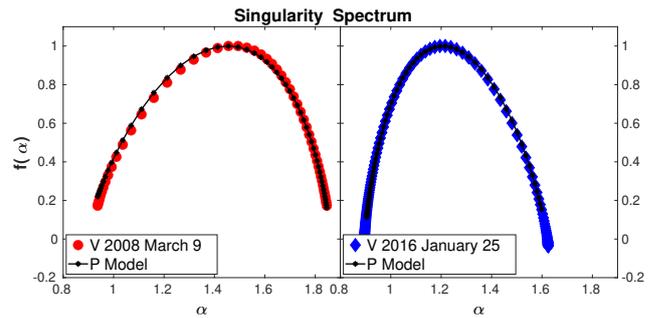

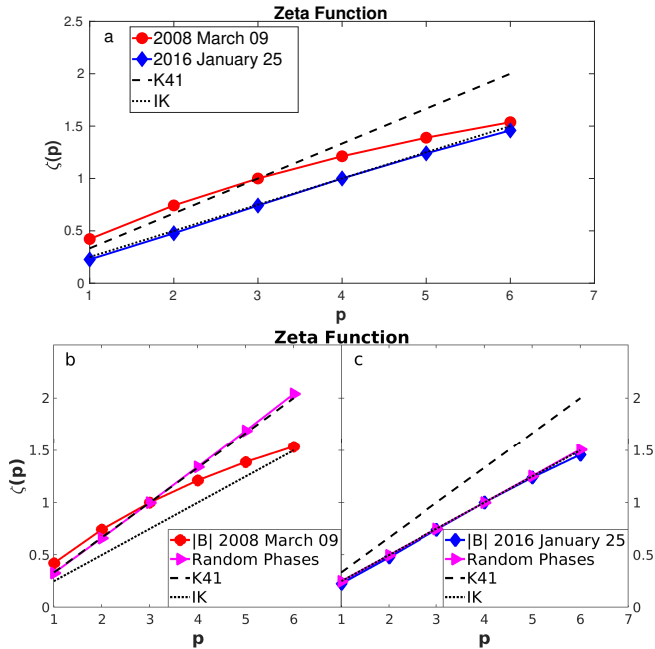

**Figure 14.** (a) Zeta functions for the magnetic field time series for 2008 March 9 (red circles) and 2016 January 25 (blue diamonds). (b) Zeta functions for |B| 2008 March 9 (red circles) and its Random Phases (magenta triangles). (c) Zeta functions for |B| 2016 January 25 (blue diamonds) and its Random Phases (magenta triangles). The dashed lines represent the K41 scaling and the dotted lines, the IK scaling.

## 6 *P*−MODEL

In section 4, we showed that the multifractal spectra of the volatility of the solar wind are predominantly due to nonlinear and linear correlations in the time series of January 25 and due to PDF and nonlinear correlations for the March 9 time series. The presence of long-range nonlinear correlations in both series is the signature of a nonlinear dynamical system (possibly with some stochastic component) governing the behavior of both series. In the present section, we employ the $p-$model (Halsey et al. 1986; Meneveau & Sreenivasan 1987) to show that both the correlations and the extreme events mentioned above are actually a consequence of turbulent energy-cascade processes with different scaling laws that depend on the presence or absence of current sheets in the original time series.

The $p-$model is a model for non-homogeneous energy-cascading process in the inertial range of fully-developed turbulence based on the generalized Cantor set. Consider that the flux of kinetic energy from eddies of size $L$ to smaller eddies is represented by a dissipation $E_L$. In the one-dimensional version of the $p-$model, $L$ is the length of an interval. Suppose that an eddy of size $L$ is unequally di-vided into two smaller eddies (i.e., two sub-intervals) of sizes $l_1 L$ and $l_2 L$, where $0 < l_1 < l_2 < 1$ are the size factors, with the energy flux $E_L$ being distributed unto these sub-eddies with different probabilities $p_1$ and $p_2$, i.e., the new dissipation values are $p_1 E_L$ and $p_2 E_L$. In practice, one can start the process with $L = E_L = 1$. Then, each new eddy is further sub-divided into two smaller eddies with the same size factors $l_1$ and $l_2$ and probabilities $p_1$ and $p_2$. This process may be repeated until the sub-intervals reach the Kolmogorov dissipation scale. At each cascading step $n$, there will be $\binom{n}{m}$ segments with length $l_1^m l_2^{n-m} L$ and dissipation $p_1^m p_2^{n-m} E_L$, for $m = 0, 1, \ldots, n$. As shown by Halsey et al. (1986) for the general two-scale Cantor set, it is possible to obtain the analytic expressions for the singularity exponent $\alpha$ and the singularity spectrum $f$ as

$$\alpha = \frac{\ln p_1 + (n/m - 1) \ln p_2}{\ln l_1 + (n/m - 1) \ln l_2}, \quad (14)$$

$$f = \frac{(n/m - 1) \ln(n/m - 1) - (n/m) \ln(n/m)}{\ln l_1 + (n/m - 1) \ln l_2}. \quad (15)$$

For each $n$ and given values of $l_1, l_2, p_1$ and $p_2$, the variation of $m$ will provide the different values of $\alpha$ and $f$ for the singularity spectrum. Since $0 \leq m \leq n$ and $m$ is an integer, larger values of $n$ provide a better definition of the spectrum. For a cascading process with direct energy dissipation in the inertial range, we have $p_1 + p_2 < 1$ (Meneveau & Sreenivasan 1987). This means that a new $dp$ dissipation parameter must be included, where $dp = 1 - p_1 - p_2$. Thus, we define $p_2 = 1 - p_1 - dp$, as well as $l_2 = 1 - l_1$, in Eqs. (14) and (15).

Figure 15 shows the MF-DFA multifractal spectra for the volatility series of March 9 (red circles) and January 25 (blue diamonds). The $p-$model fits obtained from Eqs. (14) and (15) are also shown (black line with dots). The values of $p_1, dp$ and $l_1$ were obtained with a Monte Carlo method that minimized the mean squared error between the original and fitted spectra. For March 9th, we obtained $p_1 = 0.71$, $dp = 0.17$ and $l_1 = 0.68$. For January 25th, we obtained $p_1 = 0.51$, $dp = 0.11$ and $l_1 = 0.66$. The agreement between the observational and theoretical curves confirms that the solar wind multifractal spectra can be obtained from a turbulence cascade process. This is a remarkable result, since the $p-$model was specifically elaborated to represent turbulent cascade processes, and will usually not be able to approximate the spectra of other processes.

Next, we compare the turbulent time series behind the $p-$model spectra with the observational solar wind volatility time series in terms of their PSDs. To obtain the $p$−model PSDs, we use the probabilities and size factors previously obtained with the Monte Carlo method. By iterating the generalized two-scale Cantor set model, we produce two $p$−model time series. Figure 16 shows a comparison of the solar wind volatility time series with the $p$−model time series. The two upper panels depict the solar wind series for March 9 (a) and the corresponding $p$−model (b); the two lower panels depict the solar wind series for January 25 (a) and the corresponding $p$−model (b). The qualitative similarity between observational and $p$−model time series is apparent in both cases.

**Figure 15.** Left: Multifractal spectrum for the volatility of 2008 March 9 (red circle) and its $p-$model fit (black line with dots). Right: Multifractal spectrum for the volatility of 2016 January 25 (blue diamond) and its $p-$model fit (black line with dots).

A comparison of observed and simulated PSDs is shown in Fig. 17. Figure 17(upper panels) shows the PSDs for the volatility time series of 2008 March 9 (left) and 2016 January 25 (right). The blue region between the vertical dashed lines represents the inertial range and the red line is the linear regression with slope $-5/3$ for the March 9 series and $-3/2$ for the January 25 series. Thus, the highly intermittent series of March 9 (with current sheets) exhibits a K41 scaling, whereas the January 25 series (without current sheets) shows an IK scaling. This fact had been previously established by Li et al. (2011) and confirmed by Gomes et al. (2019) using PSDs computed from the time series of $|B|$. The PSDs computed from the $p$−model time series are shown in Fig. 17(lower panels), and they reveal K41 scaling for the March 9 series and IK scaling for the January 25 series, just like in the original solar wind series. Note that in both cases the inertial range can be extended almost throughout the whole PSDs shown, since our $p$-model has small dissipation. We conclude that a K41 intermittent turbulence cascade is behind the multifractality of the current sheet-filled time series of March 9 and an IK turbulence cascade is the origin of the multifractality of the January 25 series. This result is consistent with other time series analysed by us, that show that current sheets are responsible for the K41 turbulence multifractality and the absence of current sheets results in an IK turbulence multifractality in the solar wind (see Table 4 in Gomes et al. (2019)).

## 7 CONCLUSIONS

We have presented a new methodology for multifractal analysis of solar wind magnetic field data, based on MF-DFA, volatility and surrogate time series. The MF-DFA provides a standard way to generate the singularity spectrum and the Renyi exponent; the volatility enhances the extreme events, stressing the differences between series with current sheets and series without current sheets; the surrogate time series provide a way to infer the origin of multifractality. Additionally, the $p$-model was used to reproduce the multifractal behavior of the solar wind series, indicating that a nonlinear turbulence energy cascade dynamical system is behind the observed dynamics. A similar framework for multifractal analysis, but without the volatility and the $p$-model, was used by Chattopadhyay et al. (2018) in the analysis of CME linear speed data in the solar wind. In order to keep the paper reasonably short, we have limited our presentation to only two time-series, but we have tested our techniques in other series and found that the conclusions presented are robust. An example of analysis with two other time series is included in the supplementary material (online). Further exploration of the methodology is left for future works.

Just like in Gomes et al. (2019), we found the volatility to be very useful to highlight the role of current sheets. In our case, they increase the signature of multifractality due to PDF in the singularity spectra. The surrogate analysis of both original and volatility series shows that for time series with current sheets, multifractality is due to both intermittency and nonlinear correlations; for time series without current sheets, it is predominantly produced by the long-range correlations. The $p$−model analysis reveals that those are mainly nonlinear correlations, since the process behind the statistics is a nonlinear turbulent energy cascade. So, turbulence is the common source of the multifractality, but current sheets are the source of the left asymmetry of the singularity spectrum, as well as the nonlinear scaling exponent for the structure functions. In the absence of current sheets, the small-amplitude fluctuations are the main source of the right asymmetry of the singularity spectrum. It is important to stress that despite being a multifractal process, the current sheet-free series exhibits an almost linear scaling exponent for the structure functions, which is sometimes confused with a monofractal process in the literature. Our results indicate that the Renyi exponent is more sensitive to multifractality due to correlations than the structure function scaling exponent (zeta function).

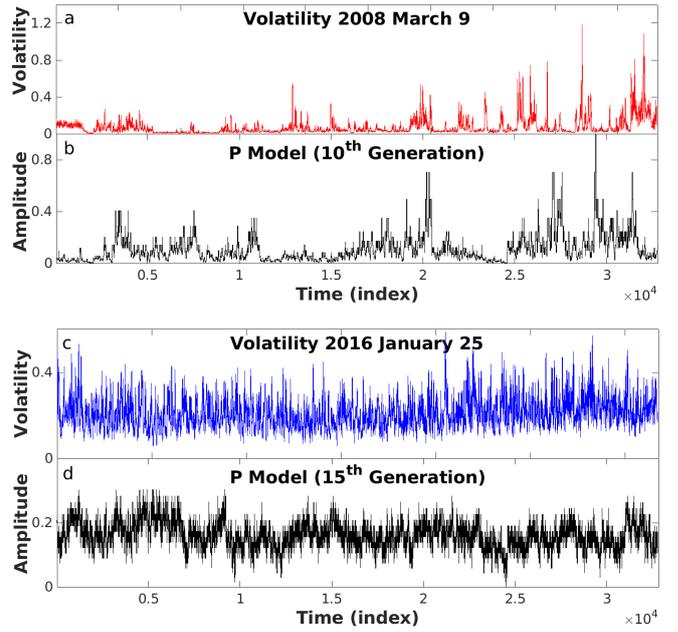

**Figure 16.** (a) Volatility time series for 2008 March 9 (red) and (b) generated $p$−model time series (black) by $10^{th}$ interation. (c) Volatility time series for 2016 January 25 (blue) and (d) generated $p$−model time series (black) by $15^{th}$ interation.

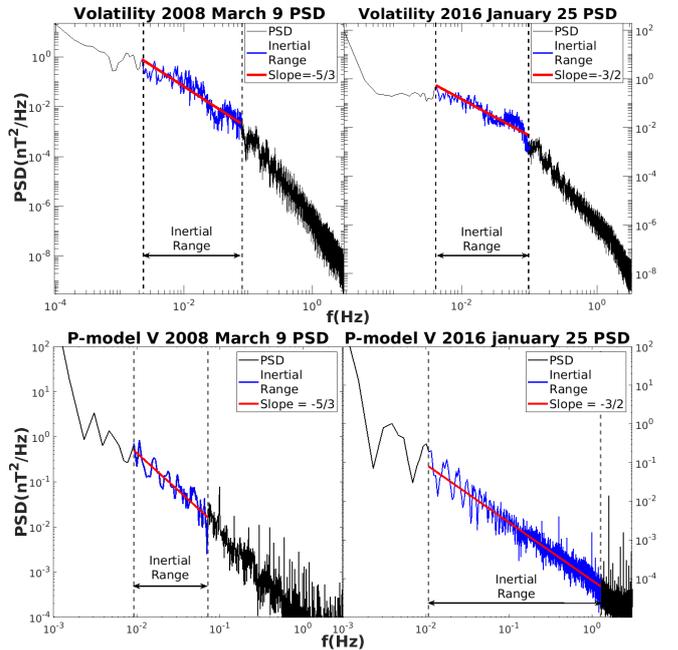

**Figure 17.** (a) Left: power spectral density for 2008 March 9 volatility. (a) Right: power spectral density for 2016 January 25 volatility. (b) Left: power spectral density for generated $p$−model time series from 2008 March 9 volatility. (b) Right: power spectral density for generated $p$−model time series from 2016 January 25 volatility. The blue regions mark the inertial range and the red lines are the linear fits for those intervals.

In dealing with separate cases where the presence or absence of current sheets is considered, we are attacking one of the "nine outstanding questions of solar wind physics", related by Vaill & Borovsky (2020), namely, the origin and evolution of the mesoscale (timescales in the range of minutes up to a few hours) plasma and magnetic-field structure of the solar wind. These current sheets have been associated with the border between adjacent flux tubes (Bruno 2019), while also being related to nonlinear turbulent interactions rather than the presence of advected pre-existing flux-tube structures (Bowen et al. 2018). In the present work, we do not focus on the origin of those coherent structures, but measure their weight on the statistics of solar wind fluctuations. We do this not only through Fourier spectral indices and the scaling of structure functions, as in Salem et al. (2009), but their contribution to multifractality is explored in depth through the MF-DFA, volatility and surrogate techniques. As we said, our results reveal that although the scaling of the structure functions may be almost linear for series without current sheets, the singularity spectra may still display broad parabolas, the signature of highly multifractal signals. Thus, the scaling exponent of structure functions is adequate to measure multifractality due to PDFs, but not for multifractality due to long-range correlations, where the Renyi exponent and singularity spectra should be adopted. Multifractal series with nearly linear behavior of the scaling exponents were also reported in Tam et al. (2010) (see their Fig. 4), where the rank-order multifractal analysis (ROMA) is employed in the description of auroral zone electric-field fluctuations.

In conclusion, the basic question related to mesoscale plasma turbulence in the solar wind is not whether it is monofractal or multifractal, but if the source of the ubiquitous multifractality is the PDF or the long-range correlations. The short answer is that in the presence of current sheets, the PDF has a strong contribution for multifractality, but in their absence, it is mainly due to correlations. It would be interesting to check if the monoscaling of the structure functions reported in previous solar wind time series, as in Kiyani et al. (2009, 2013) and Bruno (2019) for turbulence at kinetic scales, indeed reveal monofractality or if they indicate, in fact, multifractal series due to correlations and not due to intermittency.

**ACKNOWLEDGEMENTS**

L.F.G. acknowledges Brazilian agency CAPES for the financial support; E.L.R. acknowledges Brazilian agencies CAPES (grant 88887.309065/2018-01) and CNPq (Grant 306920/2020-4) for their financial support, as well as FCT—Fundação para a Ciência e a Tecnologia (Portugal); S.G. was partially supported by (i) CMUP, member of LASI, which is financed by national funds through FCT – Fundação para a Ciência e a Tecnologia, I.P., under the project with reference UIDB/00144/2020, and (ii) project SNAP NORTE-01-0145-FEDER-000085, financed by ERDF through NORTE2020 under Portugal 2020 Partnership Agreement.

**DATA AVAILABILITY**

The data used for this analysis can be obtained from European Space Agency (ESA) at the Cluster Science Archive: https://csa.esac.esa.int/csa-web/ (last access: 2 December 2020, ITA, 2020).

This paper has been typeset from a TeX/LaTeX file prepared by the author.